# Calculating the probability of detecting radio signals from alien civilizations

Marko Horvat[1]

[1]*Faculty of Electrical Engineering and Computing, University of Zagreb*
*Unska 3, HR-10000 Zagreb, Croatia*
*E-mail: marko@marko-horvat.name*

**Abstract.** *Although it might not be self-evident, it is in fact entirely possible to calculate the probability of detecting alien radio signals by understanding what types of extraterrestrial radio emissions can be expected and what properties these emissions can have. Using the Drake equation as the obvious starting point, and logically identifying and enumerating constraints of interstellar radio communications can yield the probability of detecting a genuine alien radio signal.*

## 1. Introduction

The primary goal of SETI (Search for Extraterrestrial Intelligence) projects is to receive a signal from an extraterrestrial civilization. Today, SETI projects predominantly listen to radio signals, although there are also SETI projects that search for optical [1] [2] [3] [4] and microwave [5] [6] electromagnetic frequencies.

The possibility of detecting an alien electromagnetic signal other than the radio one is not explicitly elaborated in the article but the same principles and algebra apply to all types of electromagnetic signals.

Strictly speaking, all the signals that an alien civilization could transmit can be divided into two distinct categories: a) *intentional signals* which are meant for our civilization and are deliberately sent towards the Earth; and b) *unintentional signals* which are not really meant for our civilization – they are only accidentally directed towards the Earth. The intentional signals could also be called *deliberate* and the unintentional *random* or *involuntary* signals.

A clear example of an intentional radio emission is the well-known message sent in 1973 from Arecibo radio-telescope towards the M13 globular cluster. Examples of unintentional radio emissions are much more common – they are our own radio and TV signals; and of course, all other electromagnetic emissions produced by radars, microwave communications, etc. These signals are emitted constantly, indiscriminately and equally in all directions. They are not directed towards a specific celestial coordinate and their emitting strength is determined by the mean sensitivity and distance of a "radio listener" or a "TV viewer".

As we compare these two types or categories of signals, it is obvious that unintentional signals are much, much weaker in strength than the intentional electromagnetic emissions, but they have far longer duration. In a nutshell, it can be said that intentional radio emissions are relatively strong but brief signals directed at a specific point in space, and the unintentional radio emissions are weaker but also longer lasting signals directed everywhere, toward all points in space equally. The properties of both types of signals are listed in Table 1.

It is important to understand that it is rather unlikely that intentional alien radio signals could be constantly emitted over a significantly long period of time, i.e. a few decades or centuries, because that would be an extremely power and resource consuming endeavor. More so, if such an altruistic alien civilization would want to cover a number of star systems with its intentional signal.

An evolved alien civilization that would deliberately and constantly send radio messages toward solar systems potentially harboring intelligent life is really hard to imagine, especially if the civilization discovered how to efficiently travel and communicate over interstellar distances. This would also make the altruistic civilization extremely impractical. But although such a civilization is practically improbable, it is still theoretically possible.

However, the conclusion is that if we ever were to receive a radio message from an alien civilization, it would most likely be a type of unintentional radio signal. The probability of receiving an intentional radio signal is far smaller and harder to calculate, since its parameters are, at this point in the development of our astronomy and astrobiology, dependant on subjective analysis.

Further, it should also be pointed out that radio is not the best way to communicate over large interstellar distances. It is slow, liable to transmission errors and, due to usage of relatively long frequencies, radio can't carry a lot of data. Therefore, it is logical to assume that a constantly developing civilization, such as ours, will eventually discard radio as means of communication, or at least largely minimize its use. We already know of several better methods of communication, like fiber-optic cables or directional lasers which are very difficult or entirely impossible to detect over interstellar distances. So, it is quite probable that, at some point in every civilization's development, all its radio emissions will completely stop. This will make the civilization totally radio silent and thus completely undetectable by the SETI projects.

All propositions described above can be summarized in the following conclusions:
1. If we ever receive an alien radio signal, it is unlikely that the signal will be meant specifically for us. It is far more likely that this signal will be a random alien communication.
2. There is a specific time interval during which an alien civilization uses radio communications. Before this interval, radio is beyond the civilization's technical reach, and after this interval radio will be considered obsolete. We have to be listening with our radio telescopes during this period of the alien civilization's development to receive their radio signal.
3. It is necessary to define a strictly defined volume of space in which alien radio transmissions are strong enough to be picked up. The Earth, i.e. our radio receivers, must be inside this volume of space in order to be able to detect alien signals.

All reasoning about other civilizations has to start from observations of the only alien civilization we know of – our own. Simply put, humankind is the only civilization that we know of and the only test subject upon which we can model the behavior of other civilizations that might exists in the Universe. Other paths of civilization development, concerning interstellar communications in particular, are entirely possible, but we have to ask ourselves – which is the most probable? And, the answer lies in simple logic and basic common sense.

## 2. Defining the variables

As was explained in the introduction, it is assumed that all civilizations can't and don't use radio during their entire existence. Surely, it would be very difficult to imagine a civilization that along with invention of fire and wheel also invents coil, capacitor, resistor, radio antennae and the entire science of electronics. Therefore, it should be assumed that an average time of civilization radio usage must exist. This time interval will be called $t_{comm}$. In line with this thinking, it is also quite reasonable to assume that civilizations don't last forever but that they all have certain average life span $t_{civ}$.

Naturally, $t_{comm}$ can never be larger than $t_{civ}$:

$$\begin{aligned} t_{comm} &\in R \\ t_{civ} &\in R \\ t_{comm} &\leq t_{civ} \end{aligned} \qquad (1)$$

$$0 \leq \frac{t_{comm}}{t_{civ}} \leq 1$$

This *time ratio* is very important. It will be denoted as $f_t$. The time ratio describes the probability that our civilization is searching the skies for a radio signal while an alien civilization is indeed transmitting its radio signals.

$$f_t = \frac{t_{comm}}{t_{civ}} \in [0,1] \qquad (2)$$

Due to limitations of radio communication technologies over interstellar distances – the reduction of electromagnetic signal strength is directly proportional to the square of the distance traveled – and limited sensitivity of our radio receivers, signals emanating from an alien civilization could only be detected within a certain radius around the alien radio transmitter.

Therefore, it is necessary to introduce another variable called $V_{comm}$ that describes the volume of space around the alien radio transmitter in which its signal is detectable. Obviously, if sensitivity of the receiver is better or if the signal is stronger, volume $V_{comm}$ is larger. Since radio receivers can never be 100% perfect, a technical impossibility, the receivers will never be able to receive an infinitely weak signal. The consequence is that $V_{comm}$ will always have a finite size that influences the probability of alien signal detection.

$V_{comm}$ of the intentional radio signal can greatly vary with the signal's emitting strength. It is hard to predict mean strength of a radio signal deliberately sent to us. Perhaps it might be very strong, perhaps not. Perhaps the Earth can sit right in the middle of the signal's path, perhaps not. On the other hand, $V_{comm}$ of unintentional radio signals, because they would be omni directional, depends exclusively on the mean sensitivity of our own receiving antennas and the predicted average strength of alien unintentional signals. In order to get the latter value, we can use the average emitting strength of our own TV and radio signals.

Along with $V_{comm}$ we must also consider a strictly defined volume of space, e.g. a galaxy, a cluster of galaxies, the entire Universe, in which alien civilizations could exist. This volume of space is denoted by $V_{space}$ and all civilizations that we might detect are located in it.

Since electromagnetic waves spread through 3-dimensional space in spheres, it is possible to use simple geometric formula for the volume of a sphere and describe $V_{comm}$ and $V_{space}$ by their respective radiuses $r_{comm}$ and $r_{space}$:

$$V_{comm} \in R$$
$$V_{space} \in R$$
$$V_{comm} = \frac{4}{3}\pi r_{comm}^3 \quad (3)$$
$$V_{space} = \frac{4}{3}\pi r_{space}^3$$

The mutual relationship of $V_{comm}$ and $V_{space}$ can also be determined:

$$V_{comm} \leq V_{space}$$
$$0 \leq \frac{V_{comm}}{V_{space}} \leq 1 \quad (4)$$

By substituting (4) with radiuses $r_{comm}$ and $r_{space}$ we get the following:

$$r^3_{comm} \leq r^3_{space}$$
$$0 \leq \frac{r^3_{comm}}{r^3_{space}} \leq 1 \quad (5)$$

Now, it is possible to define another important ratio, *volume ratio $f_v$*, as:

$$f_v = \frac{r^3_{comm}}{r^3_{space}} \in [0,1] \quad (6)$$

The volume ratio $f_v$ describes the probability that an alien civilization is close enough to the Earth for its signals to be detected with our radio equipment.

The volume ratio can be visualized as in Figure 1. In this figure, the Earth is shown at the center of an imaginary sphere in space, while two solar systems with alien civilizations are within our $r_{comm}$ and thus detectable. Three solar systems are outside our communications range and cannot be detected. One solar system with an alien civilization is outside of the scope of interest, i.e. further away than $r_{space}$.

As we move on with the analysis of interstellar communications between different civilizations, we must also keep in mind that at each moment in time $N_{ac}$ alien civilizations can occupy the same volume of space $V_{space}$. Theoretically speaking, we could contact any number of these civilizations: one, two, several, or even none. The value of $N_{ac}$ is calculated with the well-known Drake (or Drake-Sagan) equation [7]:

$$N_c = N_* f_p n_e f_l f_i f_c f_L \quad (7)$$

The number of currently existing alien civilizations $N_{ac}$ is simply the number of all civilizations minus one, i.e. with our human civilization omitted:

$$N_{ac} = N_c - 1 \quad (8)$$

A very important feature of $N_{ac}$ is that it is a constant. This is guaranteed by the premise of the Drake equation: the average number of civilizations in a galaxy always remains the same. If a civilization dies out and vanishes, it is immediately "replaced" by a new civilization although at an inferior level of technical development.

After we have summed up and defined all pertinent variables, we can formulate the probability function that describes the detection of radio signals from alien civilizations.

## 3. Calculating the probability

The probability that we want to calculate in this article, i.e. the probability of detecting radio signals from any alien civilization, will be simply called $p$. In order to get $p$, we must take few preliminary steps and determine several other probabilities that will eventually lead us to our final goal.

First, we must calculate the probability of receiving a signal from *exactly one* alien civilization. This probability will be denoted $p'$. It consists of two other probabilities which are calculated by mathematical ratios described in the previous section – time ratio $f_t$ and volume ratio $f_v$ – as in (9):

$$p' = f_t f_v \qquad (9)$$

In probability $p'$ we are imagining a sphere as in Figure 1 where the Earth is located at the center of the sphere and exactly one alien civilization is somewhere inside $V_{comm}$. We are also taking into account the time factor and considering only those alien civilizations that are using radio at the same time as we, the listeners, are.

The reasoning behind $p'$ can be interpreted like this: if exactly one alien civilization using radio is in our communications range, than we shall certainly have an opportunity to detect it sometime during the existence of our civilization.

Now when $p'$ is known, it is possible to find out the probability that we will *never* detect radio signals from *one* alien civilization. This probability $p''$ is the exact opposite of $p'$:

$$p'' = 1 - p' \qquad (10)$$

Using the equation above, we can calculate the probability $p'''$ that we will *never* detect signals from $N_{ac}$ alien civilizations. We simply have to repeat the stochastic event p'' $N_{ac}$ times:

$$p''' = p''^{N_{ac}} \qquad (11)$$

As can be seen, $p'''$ gets smaller when $N_{ac}$ gets bigger. This is exactly what should be expected; probability that we will never detect even a single alien civilization decreases if there are more alien civilizations to detect. The decrease is exponential. The more alien civilizations are out there, the less likely it is that we won't detect any of them.

At this point our mathematical apparatus is almost complete for the construction of the equation calculating the final probability $p$.

The probability of detecting signals from alien civilizations is equal to the probability of detecting a signal from *at least one* alien civilization. And this probability is our goal. We can detect signals from any number of civilizations, but signals from just one civilization will also "do the trick" quite nicely.

To contact *at least one* alien civilization means contacting *1, 2, 3, ... , n, n + 1* ($n$ is a whole number greater than zero) civilizations. In mathematical and logical terms, the exact opposite of *at least one* is *none*. Since (11) gives the probability of detecting *none* alien civilizations, the probability $p$ of detecting *at least one* alien civilization, is obtained by logically, and probabilistically, negating $p'''$:

$$p = 1 - p''' \qquad (12)$$

Individual occurrences of probability $p'$ (9) are independent of each other. Or, in other words, the existences of different alien civilizations are mutually independent – their existence is neither caused nor affected by each other. Because of this, $p$ has a binomial distribution [8], [9] and it is possible to construct the mathematical expression below:

$$p = 1 - \binom{N_{civ}}{N_{civ}} p'''^{N_{ac}} (1 - p''')^0 \qquad (13)$$

The equation above using binomial distribution for the reoccurrence of independent stochastic events, describes the probability that a SETI project, which scans all radio frequencies all the time, will detect a random signal from at least one and at the most, $N_{ac}$ alien civilizations.

This is almost exactly what we want to get, but there is one more step to make – or two – as is described in the next section. But first (13) has to be reduced with two simple expressions:

$$\binom{N_{ac}}{N_{ac}} = 1 \quad (14)$$
$$(1-p'')^0 = 1$$

And now, by introducing (9), (10) and (11) into (13), we get the equation which describes the probability of detecting radio signals from alien civilizations:

$$p = 1 - (1 - f_t f_v)^{N_{ac}} \quad (15)$$

(15) calculates the probability that an ideal SETI project, which scans the entire sky on all frequencies, will detect radio signals from an alien civilization. The equation can be applied for both unintentional and intentional types of signals.

## 4. Refining the probability

Two more restrictions must be introduced in (15) to make it realistic, practical and truly applicable. The first restriction concerns the number of scanned wavelengths or frequencies, and the second restriction is related to the number of extra-solar systems that a SETI project observes during its operation.

Ideally, if a SETI project observes all wavelengths that alien civilizations could use for interstellar communications, there are no restrictions or limitations. Obviously, in the search for extraterrestrial radio signals, we must listen on as many radio channels as possible – preferably, on all. But if only a part of available wavelengths is observed, as is the case with any realistic situation, then, this wavelength problem does exist and it reduces the probability in (15) This restriction is called *the wavelength ratio $f_\lambda$*:

$$f_\lambda \in [0,1] \quad (16)$$

The ratio can be calculated discretely or continuously depending on how the wavelength problem is defined. The simplest way to calculate $f_\lambda$ is to divide the length of the spectrum of all wavelengths observed by the SETI project ($\Delta\lambda_o$) with the spectrum length of all wavelengths that alien civilizations use for radio communication ($\Delta\lambda_c$):

$$f_\lambda = \frac{\Delta\lambda_o}{\Delta\lambda_c} \quad (17)$$

The discrete way of calculating $f_\lambda$ is analogous to the previous equation but, instead of the continuous length, the number of respective wavelengths is used:

$$f_\lambda = \frac{|\lambda_o|}{|\lambda_c|} \quad (18)$$

$|\lambda_o|$ is the number of all observed wavelengths and $|\lambda_c|$ the number of all wavelengths used for alien radio communications.

In both equations, the following relationships apply:

$$\Delta\lambda_o \leq \Delta\lambda_c$$
$$0 \leq \frac{\Delta\lambda_o}{\Delta\lambda_c} \leq 1$$

$$|\lambda_o| \leq |\lambda_c|$$
$$0 \leq \frac{|\lambda_o|}{|\lambda_c|} \leq 1$$

Determining the true value of $|\lambda_c|$ can be tricky since we can only make educated guesses. A very rough, but a practical estimate for $|\lambda_c|$ is 1, which takes into account only the famous wavelength of neutral hydrogen at 21cm. More precise estimates of $|\lambda_c|$ will increase its value. The same applies to $\Delta\lambda_c$.

The second limitation to the probability in (15) is in many ways similar to the wavelength ratio. As an example, it is easy to imagine that an alien radio signal, continuous and unintentional or intermittent and intentional, reaches the Earth, but we can't hear it because our radio antennas are directed elsewhere. This premise is well-known and the rationale how to counter it is built into every SETI project: "Have more radio telescopes, scan as much sky as possible and increase the likelihood of detecting alien radio signals". As a compromise to scanning every visible star in the sky (obviously a very demanding task) an educated guess can be made and only those extra-solar systems where life can thrive are scanned. Of course, to be valid, this guess has to have a reasonable degree of certainty. A perfect, but impossible, SETI project would scan the whole sky. Only then, it could be guaranteed that no signal has been overlooked.

The second limitation to (15) is *the area ratio $f_A$*:

$$f_A \in [0,1] \quad (19)$$

The area ratio can also be calculated continuously, taking into account the sky area itself, or discretely, by counting the number of observed extra-solar systems. Thinking continuously, we take the area of sky observed by a SETI project ($\Delta A_o$) and divide it with the area of the entire sky ($\Delta A_c$):

$$f_A = \frac{\Delta A_o}{\Delta A_c} \quad (20)$$

In discrete terms, we will use the number of extra-solar systems that are observed by the SETI project ($|A_o|$) and divide it with the number of solar systems where alien civilizations can exist ($|A_c|$). As with $f_\lambda$ and parameters $|\lambda_c|$ or $\Delta\lambda_c$, one is more or less in the dark when trying to establish the true value of $|A_c|$.

$$f_A = \frac{|A_o|}{|A_c|} \quad (21)$$

Since both restrictions, $f_\lambda$ and $f_A$, affect all observations equally, they are applied to the joint probability of $p$ in (15), and not to the probability $p'$ of the signal detection of an individual alien civilization (9).
Finally, after applying $f_\lambda$ and $f_A$ to the (15), it realistically describes the probability of detecting radio signals from an alien civilization:

$$p = f_\lambda f_A \left(1 - (1 - f_t f_v)^{N_{ac}}\right) \quad (22)$$

(22) gives the probability that a realistic SETI project, which scans only a portion of the sky on a finite number of frequencies, will detect radio signals from an alien civilization. The equation can best be applied to unintentional, and only theoretically, to intentional type of signals.

Changing the value of either factor $f_\lambda$ or $f_A$ must be done very carefully since they have a big impact on the final value of the probability of detecting unintentional alien radio signals $p$. In order to simplify matters and maximize $p$, we can assume $f_\lambda = 1$ and $f_A = 1$. Although this represents only the best possible situation that is hardly achievable, it can be used to set the upper limit of $p$.

Perhaps the last, but certainly not the least important fact about the parameters in (22),, is that they fall into two distinct categories: variables and constants. We, the observers, can influence values of $f_\lambda$, $f_A$ and $f_v$ but we cannot modify $f_t$ and $N_{ac}$. The latter two parameters are preset and cannot be affected by observers. Therefore, it can be said that $f_\lambda$, $f_A$ and $f_v$ are variables, and $f_t$ and $N_{ac}$ are constants.

## 5. Applying the figures

Let's take a set of different values and apply them to (22). The result is in Table 2.
The table is divided into 4 sections, each with 5 cases, with the case 15 being the near-optimum case. First three sections cover variations in $t_{comm}$, $t_{civ}$, $r_{comm}$, $r_{space}$ and $N_{ac}$ parameters leading to the best case. The last section in Table 2 tests how variations in $f_\lambda$ and $f_A$ affect the best case.
The first five cases in the table take modest value for $t_{comm}$ of only 500 years, $r_{space}$ is roughly equal to the radius of our galaxy and $r_{comm}$ is a realistic $1/5^{th}$ of $r_{space}$. $N_{ac}$ is conservative, ranging between 5 and 30 alien civilizations in the galaxy while $t_{civ}$ varies between 4 million years and 500 000 years. As can be seen, these figures give a very small, almost negligible, probability $p$. The primary reason is long civilization lifespan and short usage of radio technology. Improving our detection capabilities would also improve the whole picture.
The cases marked 6 to 10 introduce more variations in figures of the first five cases. $t_{comm}$ is increased to 1000 years, ratio of $r_{comm}$ and $r_{space}$ is ½ and it is assumed that the number of alien civilizations in our galaxy is larger, varying from 60 to 140. However, these cases also yield a very low probability of detecting alien signals $p$. A small $t_{comm}$ / $t_{civ}$ ratio has the greatest negative impact and is the main reason of low $p$ which, even increased value of $N_{ac}$, cannot overcome. Only the 10th case has a noticeable $p$ but that dictates a very high number of alien civilizations in the galaxy (140). Since they have a lifespan of 500 000 years, it means that a great number of civilizations would have to be approximately at the same development level – an interesting implication.

The third set of cases (11-14) assumes relatively high value of 2500 years for $t_{comm}$ and short value of 250000 years for $t_{civ}$. The ratio $r_{comm}$ / $r_{space}$ is perfect – 1.0. $N_{ac}$ is reasonable, ranging between 10 and 100. This set of cases finally has a significant $p$: ~ 10% - 65%. Only this set represents successful SETI projects with tangible success factor. It is good to know what are the prerequisites of such SETI projects.

The 15th case should represent a desirable optimum: 300 alien civilizations, our receivers are perfect, $t_{civ}$ is short and $t_{comm}$ is long. This case gives a near 100% success rate. This is certainly very appealing from the standpoint of any SETI project. But the most interesting thing demonstrated in the 15th case is the exponential dependence of $p$ to $N_{ac}$: $N_{ac}$ = 100 gives $p$ = 63.397%; $N_{ac}$ = 150, $p$ = 77.855%; $N_{ac}$ = 200, $p$ = 86.602%; $N_{ac}$ = 250, $p$ = 91.894%, $N_{ac}$ = 300, $p$ = 95.096%; $N_{ac}$ = 350, $p$ = 97.033%; and $N_{ac}$ = 400, $p$ = 98.205%. With 200 or 250 alien civilizations using radio in our communication range, we can be almost completely certain that we will pick up at least one of them.

Finally, the last set (cases 16-20) works with the best cases (11-15) and experiments with different values of $f_\lambda$ and $f_A$ in an attempt to give them a realistic outline. The last set of cases looks promising: the lowest probability is about 0.3% while the highest is over 50% (53.491%)

The result in Table 2 can be broadly commented on like this: the probability of detecting alien civilizations by SETI projects can be either very small or very high. Minute variations in the equation's parameters have a great influence on the final probability.

The highest probability of contact is achieved in the scenarios with many alien civilizations that have comparably short life spans and long usage of radio communications. At the same time, our listening techniques have to be near perfect.

The reasons behind this optimum probability are understandable; compared with their lifetime, civilizations always have relatively short radio-communication status. If there are a lot of them, in other words – if they die rather quickly and are replaced with new civilizations, then, at every point in time, there is a fair number of them that are using the radio. Long-lived civilizations pass their radio-communication status and still remain in $N_{ac}$ but are undetectable by radio. In some respect, they are occupying a place in $N_{ac}$ where some younger and radio active civilization might sit.

It is somewhat paradoxical that we should be opportunistic for short-lived alien civilizations, when at the same time, we want to learn more about them.

## 6. Conclusion

When knowing the probability of contacting aliens with our radio telescopes, one very logical question immediately pops up: when will it happen? In order to answer this important question, we have to look at the only time factor in the whole equation – $f_t$. This ratio cannot yield the exact time when the event would occur because it describes only the overlapping, or synchronization, of radio usage by our two civilizations. There is no moment in time embedded in the ratio. Therefore, it becomes obvious that the exact date of contacting aliens remains unknown. We may contact them tomorrow, maybe in a hundred years, or maybe it will take even longer. Furthermore, the time ratio $f_t$ represents an average value. It varies from one civilization to another, and with our current knowledge of astrobiology, we can only speculate about its exact value. In fact, it would be reasonable to assume that $f_t$ follows Gaussian distribution.

The probability of radio-contacting alien civilizations consists of two disjunctive subsets: the probability of receiving intentional alien signals and the probability of receiving unintentional alien signals. The detection probability of intentional signals is assumed to be negligible or zero, and this article focused only on the unintentional signals. Since unintentional signals are emitted non-stop, and not only during 73 seconds, like the intentional Arecibo radio message, another major and unpredictable time constraint defining the message's duration could have been taken out of the final (22). At this point in our development, we have to concentrate our efforts on detecting the involuntary alien emissions, e.g. the alien TV and radio. Only with these types of signals, it is realistically possible to make an educated guess about their detection.

As our science, especially astronomy and astrobiology, undoubtedly evolves, we shall certainly be capable to understand better all the factors behind the probability of detecting radio signals from alien civilizations. We shall be able to refine the probability, improve it, and finally build upon it to gain more knowledge about our galactic surroundings and the Universe we are immersed in.

## 8. Tables

| Attributes | | Signal type | |
|---|---|---|---|
| | | Unintentional | Intentional |
| | Mean strength | Low | Low/Medium/High |
| | Directed | No | Yes |
| | Containing specific information | No | Yes |
| | Duration | Long | Short |
| | Detection possibility | Low/Medium/High | Low |

**Table 1 – Comparing types of alien signals**

| | 1 | 2 | 3 | 4 | 5 | 6 | 7 | 8 | 9 | 10 |
|---|---|---|---|---|---|---|---|---|---|---|
| $t_{comm}$ | 500 | 500 | 500 | 500 | 500 | 1.000 | 1.000 | 1.000 | 1.000 | 1.000 |
| $t_{civ}$ | 4.000.000 | 2.000.000 | 1.000.000 | 750.000 | 500.000 | 4.000.000 | 2.000.000 | 1.000.000 | 750.000 | 500.000 |
| $r_{comm}$ | 10.000 | 10.000 | 10.000 | 10.000 | 10.000 | 25.000 | 25.000 | 25.000 | 25.000 | 25.000 |
| $r_{space}$ | 50.000 | 50.000 | 50.000 | 50.000 | 50.000 | 50.000 | 50.000 | 50.000 | 50.000 | 50.000 |
| $N_{ac}$ | 5 | 10 | 15 | 25 | 30 | 60 | 80 | 100 | 120 | 140 |
| $f_\lambda$ | 1,000 | 1,000 | 1,000 | 1,000 | 1,000 | 1,000 | 1,000 | 1,000 | 1,000 | 1,000 |
| $f_A$ | 1,000 | 1,000 | 1,000 | 1,000 | 1,000 | 1,000 | 1,000 | 1,000 | 1,000 | 1,000 |
| p | 0,012% | 0,050% | 0,150% | 0,333% | 0,598% | 0,747% | 1,980% | 4,878% | 7,691% | 13,070% |

| | 11 | 12 | 13 | 14 | 15 | 16 | 17 | 18 | 19 | 20 |
|---|---|---|---|---|---|---|---|---|---|---|
| $t_{comm}$ | 2.500 | 2.500 | 2.500 | 2.500 | 2.500 | 2.500 | 2.500 | 2.500 | 2.500 | 2.500 |
| $t_{civ}$ | 250.000 | 250.000 | 250.000 | 250.000 | 250.000 | 250.000 | 250.000 | 250.000 | 250.000 | 250.000 |
| $r_{comm}$ | 50.000 | 50.000 | 50.000 | 50.000 | 50.000 | 50.000 | 50.000 | 50.000 | 50.000 | 50.000 |
| $r_{space}$ | 50.000 | 50.000 | 50.000 | 50.000 | 50.000 | 50.000 | 50.000 | 50.000 | 50.000 | 50.000 |
| $N_{ac}$ | 10 | 40 | 80 | 100 | 300 | 10 | 40 | 80 | 100 | 300 |
| $f_\lambda$ | 1,000 | 1,000 | 1,000 | 1,000 | 1,000 | 0,250 | 0,500 | 0,750 | 0,750 | 0,750 |
| $f_A$ | 1,000 | 1,000 | 1,000 | 1,000 | 1,000 | 0,125 | 0,250 | 0,250 | 0,500 | 0,750 |
| p | 9,562% | 33,103% | 55,248% | 63,397% | 95,096% | 0,299% | 4,138% | 10,359% | 23,774% | 53,491% |

**Table 2 – Probability of detecting radio signals from alien civilizations**

## 9. Figures

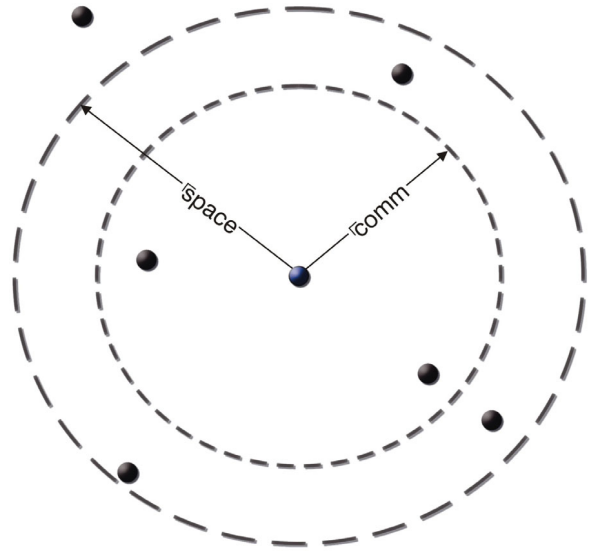

**Figure 1 – Two-dimensional visualization of the volume ratio**